# Acquisition of a Project-specific Process

Olga Jaufman[1], Jürgen Münch[2]

[1]DaimlerChrysler AG,
89081 Ulm, Germany
Olga.Jaufman@daimlerchrysler.com

[2]Fraunhofer Institute for Experimental Software Engineering (IESE)
Sauerwiesen 6, 67661 Kaiserslautern, Germany
Juergen.Muench@iese.fraunhofer.de

**Abstract.** Currently, proposed development processes are often considered too generic for operational use. This often leads to a misunderstanding of the project-specific processes and its refuse. One reason for non-appropriate project-specific processes is insufficient support for the tailoring of generic processes to project characteristics and context constraints. To tackle this problem, we propose a method for the acquisition of a project-specific process. This method uses a domain-specific process line for top-down process tailoring and supports bottom-up refinement of the defined generic process based on tracking process activities. The expected advantage of the method is tailoring efficiency gained by usage of a process line and higher process adherence gained by bottom-up adaptation of the process. The work described was conducted in the automotive domain. This article presents an overview of the so-called Emergent Process Acquisition method (EPAc) and sketches an initial validation study.

## 1 Introduction

Nowadays, automotive products are becoming more and more complex. In order to ensure the quality of safety critical products like vehicles, effective and efficient development processes are needed. As projects have different contexts and goals, tailoring methods are needed that allow adapting the generic processes to the project-specific needs. The tailoring approaches used in practice (e.g., the tailoring approach proposed by the V model [10]) usually involve checking conditions and removing objects of the base model. The V model distinguishes between tailoring at the start of a project and tailoring in the course of the project at defined points in time. One difficulty of such tailoring is the identification of the regression process modification to be performed. For example, a change of four product artifacts can result in further changes of 26 process models [12]. Further, more process tailoring often requires not only the removal of process objects, but also their replacement, or the addition of new objects. The V model tailoring method does not define how to deal with such kinds of process modifications.

To tackle the problem, different tailoring approaches are proposed in the literature. These tailoring approaches can be classified into two types [12]: component-based approaches and generator approaches. The component-based approaches try to build a project-specific process based on the process parts. The generator approaches try to build a project-specific process by instantiating a typical process architecture. The advantage of component-based approaches is the ability to support reuse of process fragments (e.g., processes gained by descriptive process modeling). The main deficiency of component-based approaches is the lack of support for process adaptation and for guaranteeing consistency. The advantage of generic approaches is their ability to assure consistency and to reuse process fragments. The disadvantage of the generic approaches is the lack of support for process fragment reuse.

Our proposed solution to the problem is the Emergent Process Acquisition (EPAc) method. This method uses a domain-specific process line for top-down tailoring and refines the tailored process based on the process activities performed in a first process iteration. In this way, the initial variant of the emergent process is built. An *emergent process* is a process that needs to cope with changing goals and context characteristics, which can only be anticipated to a very limited extend before the start of the project. Therefore, the process itself needs to be highly adaptable, and support for the adaptation is necessary.

Typical reasons for the need for emergent processes are:
- Changing requirements. The requirements are not completely known at the start of the project and, in addition, the effects of new or modified requirements on the development process cannot be anticipated. Thus, the activities to be performed can only be detected in the course of the project, too.
- Changes in the project environment. One example for a business environment change is the establishment of a new business relationship (e.g., a new international collaboration). One example for a change in the development environment is a replacement of a validation technique (e.g., a project team follows a prescriptive process and recognizes that the process is not really efficient to perform module testing).

The expected advantage of our method is higher process acceptance by project teams, as the process is based on experience from past projects and feedback from actual project performance.

The paper is structured as follows: The second section describes the background information. The third section describes the EPAc method. The fourth section briefly sketches our experience gained with the usage of the EPAc method. The fifth section discusses related work and strengths of our EPAc method. Finally, Section 6 gives a short summary and an outlook on future work.

## 2 Background Information

A systematical state-of-the-practice analysis performed by DaimlerChrysler [6] resulted in the awareness that the software development processes are too generic for operational use. The applied tailoring approach [6] does not provide enough support to project teams. This has two reasons: First, it is difficult for process engineers (who are usually also playing a role in a development team) to identify the regression process modification if the process changes. Second process tailoring often requires not only removal of process objects, but also the replacement or addition of new objects. The applied tailoring method does not define how to deal with such kinds of process modifications. Thus, a method for acquiring a project-specific process is needed, which helps project teams to tailor their prescriptive process to their project-specific needs.

## 3 Acquisition Method

Our acquisition method consists of two main steps. In the first step, a domain-specific process line is used for top-down tailoring at the start of a project. The purpose of the process line is to provide domain knowledge necessary to define a suitable software development process. The approach on how the process line is built and the schema of the process line can be found in [8]. After the first development cycle, the top-down tailored process is refined based on the tracked process data. This two-step tailoring allows reducing the deficiencies of traditional tailoring methods. The next section describes the tailoring method in more details.

The acquisition method consists of four main steps (see Figure 1).

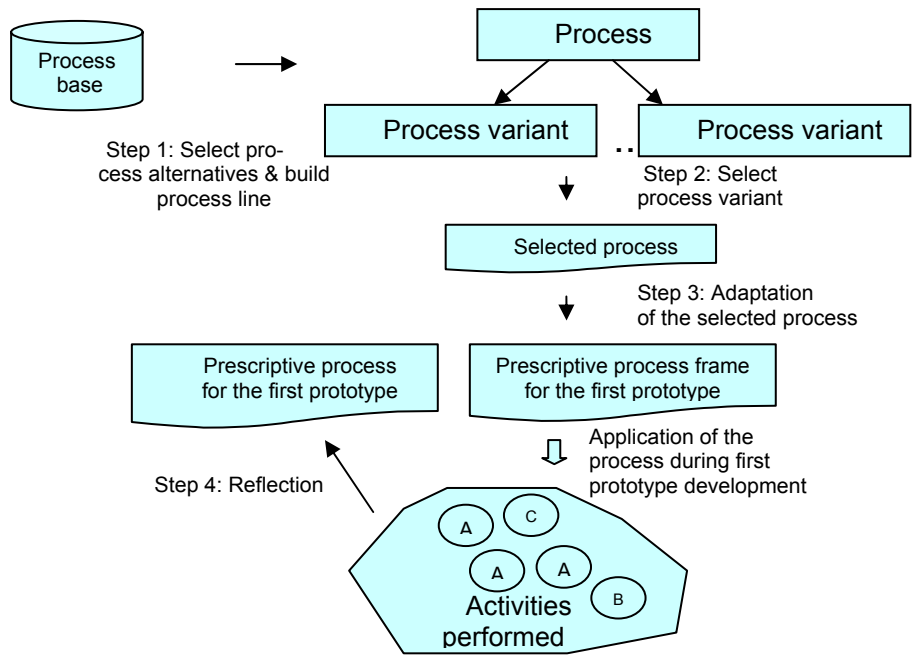

**Fig. 1: Method for Acquisition of a Project-specific Process**

In the first step, initial suitable process variants are selected from the process base based on the project characteristics. Second, the process line is built for the selected processes. The notion behind the process line is to capture the commonalities in reusable process building blocks and to construct the explicit process variants based on the process deviations, and by reusing the individual building blocks as applicable. In the second step, the process designer iteratively selects a process variant from the process line. Then the selected process variant is adapted to the project by removing unneeded process objects and adding the missing process objects. In this way, the prescriptive process for the development of the first prototype is built. Finally, the fourth step involves the elicitation of activities performed during development of the first prototype and refinement of the prescriptive process based on the tracked process objects. The following sections describe the four steps in more details.

### 3.1 Process Selection

The selection of a process variant from the process line consists of three main steps (see Figure 2). Each step is described by the attributes goal, input, activities, and output.

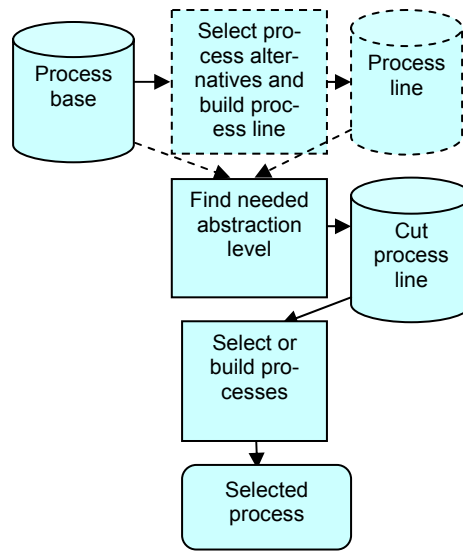

**Fig. 2: Technique for Process Selection**

**Step 1: Select process alternatives and build process line**
*Goal:* The activity is performed in order to find the most suitable process variants from the process base and built the common core of the processes. In this way, the process core contains the most important process activities.
*Inputs*: Process base, project characteristics, the importance of the project characteristics, the number of desired process variants.

*Activities*:
 1. Priorize the project characteristics.

2. Give number of desired process variants.
3. Select the most suitable process variants from the process base.
4. Build process line for selected process variants. The method for building the process line is described in [8].

The third and fourth activities are supported by a tool.
*Output*: The process line.

**Step 2: Find needed abstraction level**
*Goal:* This step has to be performed in order to find the needed process description abstraction level for the project team. The idea behind this step is that the fewer experience the project team members have, the more detailed process description they need.
*Input:* Process line, where each process variant has an abstraction index
*Activities:* Starting from the first abstraction level (i.e., with the domain-specific processes), the process designer looks at the process core of abstraction level i. If the process is too abstract from the designer's point of view, then the process designer can navigate through the process line by knowing the semantic of the abstraction index of the process variants. The tool gives the process variant with the desired abstraction index. If desired, it is possible to get the parts of the selected process variant with different abstraction levels. The process designer navigates till he finds the needed abstraction level or until the process line does not provide any more detailed process descriptions. If the process designer can find a process with the needed abstraction level, then a Cut_process_line is built by assigning variants of the process line that have the selected abstraction level. If no detailed process can be found, then the process with the most suitable abstraction level is selected.
*Output:* The process line, which contains the processes with the needed or most suitable abstraction level.

**Step 3: Select the process interactively**
*Goal:* This step can be performed in order to find the specifics of the provided process variants and to use them as input for adaptation.
*Input*: Cut_process_line
*Activities*: First, if Cut_process_line contains more than one process variant, then the supporting tool shows the variants of the cut_process_line with explicit marking of the differences between the process variants with respect to the process core. After the process designer selects one of the process variants, the tool marks the process as the selected process. If it is desired to see the difference between the selected variant and other process variants in the cut process line, the tool shows these. If the process designer would like to select another process variant, the tool provides the possibility to do this.
*Output*: The selected process variant.

## 3.2 Process Adaptation

The process adaptation can be performed in two ways (see Figure 3). If the effort for the process adaptation is lower than the effort for building a new process variant (i.e., ROI > 1), then the selected process should be adapted, otherwise, a new process should be built.

**Step 1: Adapt meta model**
*Goal:* This step is to be performed in order to tailor the process attributes to the project context.
*Input*: Selected process, meta model (which consists of the following process attributes: process phases, phase pre-conditions, phase post-conditions, delivery time, maturity of deliverables, activities, activity pre-conditions, activity post-conditions, priority of the activity, inputs needed to perform the activities, outputs needed to perform the activities, the interfaces to support processes, roles performing the activities)

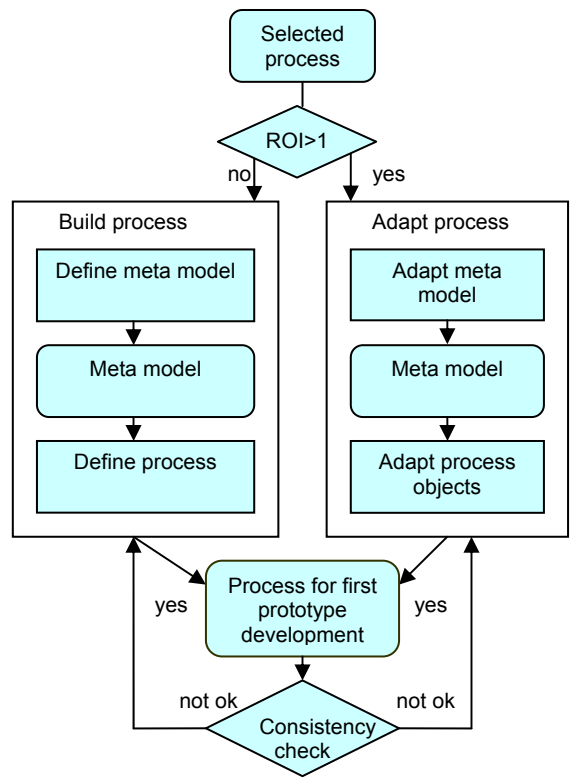

**Fig. 3: Technique for Process Adaptation**

*Activities*: The tool shows the process attributes that constitute the meta model and allows the process designer to delete unnecessary process attributes and add missing ones.

*Output*: The process meta model.

**Step 2: Adapt process instance**
*Goal:* This step has to be performed in order to tailor the process objects to the project context.
*Input*: meta process, selected process, if existent (1) to-do list, which describes the activities performed by project team members, (2) information about communication within the team, (3) data dependencies (e.g., between two parameters of a control device or between two control devices)
*Activities*: If any data (see (1)-(2)) exists, the tool shows the discrepancy between the selected process and available data. Based on this discrepancy, the process designer has the possibility to tailor the process objects by following the standard approach or the user-defined tailoring.
Standard tailoring process:
1. If the meta model contains the process attribute "milestone", then
    a) remove unnecessary milestones
    b) add missing milestones
   Output: a process with an adapted milestone
2. If the meta model contains the process attribute "phases", then the tool iteratively shows these phases for each milestone in the milestone set that are to be performed in order to achieve the milestone. Thus, for each milestone the process designer can
    a) remove unnecessary phases and
    b) add missing phases.
3. For residual process attributes (composing the meta model), the unnecessary objects are removed and missing objects are added in a similar way.

User defined process:
   The tool shows the process attributes of the meta model and allows the process designer to select a process attribute for modification. For each selected process attribute, the tool allows the process designer to remove unnecessary objects and add the necessary objects.
*Output*: the process for the first prototype development.
*Comment*: A process object cannot be removed without the approval of the project manager, if it has the priority "minimal requirement".

### 3.2.1 Build process
The process construction consists of two main steps: define meta model and define process for the meta model. In the following, these steps are described in more details.

**Step 1: Define meta model**
*Goal*: This step has to be performed in order to define milestones.
*Input:* Customer requirements on the product, selected process, if existent, project plans from past similar projects
*Activities:*

The project manager first defines the project goals based on customer requirements. Based on the project goals and similar project plans, the project manager defines milestones. These include (1) artifacts to be delivered, (2) the time point in the project for delivering the artifacts, (3) the maturity of the delivered artifact, (4) the persons responsible for artifact delivery. The milestones are captured by the tool. Furthermore, the project manager defines the process attributes with respect to which the project-specific process should be described (e.g., tasks to be performed, priority of the task).
Output: meta model

**Step 2: Define process**
*Goal*: This step has to be performed in order to define the objects for the process attributes of the meta model and the relationships between the objects.
*Input*: milestones, meta model, if existent (1) to-do list, which describes the activities performed by project team members (e.g., log list), (2) information about persons performing the tasks in the to do list, (3) data dependencies
*Activities*:
If there is a to-do list, that describes the activities performed by project team members to achieve similar milestones in past projects, the quality manager defines activities to be performed based on this list. For each activity, the quality manager creates a task. This is done by selection of the milestone (see previous step), which describes the context in which the task should be performed. Additionally, for each task the process attributes defined in the meta model are defined. For example, the priority of the task, the person responsible for the task execution, the time in which the task should be performed, can all be defined. Furthermore the tool allows its users to refine the tasks into sub-tasks, to delegate the tasks to other persons, or to inform the needed persons. The persons who need to be informed are identified by the tool based on the dependencies between the data (e.g., between two parameters of a control device or between two control devices), if such are known. If any information is missing, then the needed process attributes have to be defined based on one's own implicit experience.
*Output*: process for the first prototype development.

**3.2.2 Check Consistency**
*Goal*: This step has to be performed in order to prove the process stringency. This is important to ensure the required product quality.
*Input:* Process for the first prototype development, selected process
*Activities:*
The project manager checks that each process milestone and activity from the selected process with the priority "minimal requirement" is available in the process for the first prototype development. If a milestone or an activity is missing and is needed from the project manager's point of view, then the tool allows the project manager to add the missing milestones and activities.
Output: adapted process for the first prototype development.

### 3.3 Process Reflection

The process reflection consists of two main steps: first, elicitation and analysis of performed process objects and second, refinement and adaptation of the prescriptive process. In the following, these two steps are described in more details.

**Step 1: Elicit and analyze performed process**
*Goal*: This step has to be performed in order to understand the process actually performed by project team members, and to identify the delta between prescriptive and performed process.
*Input:* the process for the first prototype development
*Activities:* The InStep tool [4] supporting the coordination between project team members logs the status of activities performed and the time when the activities are performed. As output, the tool delivers a text log file. Our converter tool produces a xml file from the text file. This xml file is used as input for the ProM [1] and InterPol [9] tools. The ProM tool elicits a model of the performed process. The InterPol tool performs the delta analysis.
*Output*: performed process and the delta.

**Step 2: Refine prescriptive process**
*Goal*: This step has to be performed in order to update the prescriptive process based on elicited process objects.
*Input:* the process for the development of the first prototype (=prescriptive process), performed process, delta between performed and prescriptive process
*Activities:* The tool allows the project manager to add the missing process objects to the prescriptive process and to delete the unnecessary project objects. When deleting the milestones and activities with the priority "minimal requirements", the tool asks the project manager whether he/she really wants to remove the process object with the priority "minimal requirements". When deleting a process object with the priority "minimal requirement", the tool requires justification of this deletion. Here, an algorithm described in [13] is used for on-the-fly adaptation of the prescriptive process.
*Output*: refined prescriptive process, which is to be understood as a project specific process.

## 4 Validations and Gained Experience

We validated the instantiation method in the context of a initial study. In the following, the study definition, design, and the results are briefly described.

### 4.1 Study Definition and Planning

In the context of the study, we compared the Emergent Process Acquisition (EPAc) method with the tailoring method proposed by the V model [10], as the V model is widely used for system development in practice. The goal of the study was to evalu-ate the effect of the EPAc method on the effort to develop a bus control system and on the quality of the developed system.

The hypotheses in the study were:
1. The effort for developing a system according to the V model is higher or equal to the effort for developing the system according to the emergent process designed by using the EPT method.
2. Product quality developed by following the V model is higher or equal to the product quality developed by following the emergent process designed by using EPT method.
3. The satisfaction of the project team using the EPAc method is higher than the satisfaction of the project team following the tailored V model.

**4.2 Study Design and Operation**

The study was designed as follows:

The factor (i.e., the independent variable) is the development process followed by students. The treatments (i.e., particular values) of the factor are (1) the process designed by tailoring the V model and (2) the process designed by using the EPT method.

The main difference between the development following the V model and the emergent process is the process stringency. The emergent process provides more flexibility at the start of a project and becomes more stringent during the course of the project. The tailored V process has the same stringency during the course of the whole project. So the groups following the emergent process are able to better reflect on their development experience than the group following the V model, as the emergent process provides more flexibility at the start of the project than the V tailored process. Following the design principle of "balancing", we balanced the number of students in the groups. Thus, each group consisted of nine students. The students were randomly assigned to the groups, fulfilling the design principle of "randomizing". Furthermore, in order to ensure tcomparability of the study results, the following independent variables had the same treatments:

- *Customer experience*: is selected as independent variable, because a customer with little experience may state system requirements that are too ambiguous. To avoid the effect of this variable, the same person stated and clarified the requirements to all three groups.
- *Complexity of developed product*: is selected as independent variable, since the more complex the product, the higher the development effort and the higher the probability for development faults. To avoid the effect of this variable, all three groups had to develop the same system for control of bus doors and lights.
- *Environment dynamism:* We simulated the same changes in the development environment (e.g., new or changed requirements, application of a new tool). This variable is considered since the frequent changes in the development environment usually cause additional development effort.

- *Tool support:* is considered since the tools can affect the development effort. Thus, all groups use the same tool chain.
- *Instrumentation:* We provided the same measurement and preparation support for all students.

The factor treatments are assigned according to the blocking design principle. We decided that two groups were to follow the emergent process and one group had to follow the V tailored process. This assignment was meant to help us interpret the study results with more significance.

Regarding our study design, we selected 28 advanced students from eights/ninth semesters, taught them the needed foundation in software engineering and tool usage, and provided them with the needed instrumentation support (e.g., the process line for the emergent groups, forms for data elicitation). After four weeks of preparation, the study started. The study duration was 14 weeks. The study consisted of three iterations. During each iteration, a prototype should be incrementally developed. Two so-called emergent groups (E1 and E2) developed the system following the emergent process and one group developed it following the tailored V model. In each emergent group, one student was selected for the role of "emergent coach". The responsibility of the emergent coach was to design a project-specific process and to manage the team. The emergent coaches designed the project-specific process by using the process line and by following the EPT method. The two emergent groups followed the project-specific process designed by their emergent coach.

Each week,
- the groups delivered the artifacts with respect to their project plan and the completed data collection forms
- they received feedback with respect to the quality of their artifacts
- a meeting between the study supervisors and the students took place. At the meeting, the students asked questions and provided feedback to the study supervisors.

The collected data was analyzed weekly. Regarding data that seemed to be unrealistic, the students were asked directly. Additionally, the summary of the collected data was reviewed by the students to avoid misunderstandings.

### 4.3 Results Analysis and Interpretation

In order to decide about the hypotheses, we derived metrics for *productivity*, *product quality*, and *project team satisfaction* by following the GQM method [15]. In order to evaluate productivity, we compared the effort spent per activity type (see Table 1). The first row of the table shows the activity types considered.

|  | 10-17.11.2004 | | | 17-22.11.2004 | | | 23-30.11.2004 | | | 01-07.12.2004 | | | Total effort | | |
|---|---|---|---|---|---|---|---|---|---|---|---|---|---|---|---|
| Aktivität | EG 1 | EG 2 | VG | EG 1 | EG 2 | VG | EG 1 | EG 2 | VG | EG 1 | EG 2 | VG | EG1 | EG2 | VG |
| Communication customer | 0 | 0 | 0 | 270 | 300 | 270 | 0 | 0 | 0 | 0 | 0 | 30 | 270 | 300 | 300 |
| Communication TG | 0 | 0 | 600 | 210 | 120 | 0 | 90 | 60 | 60 | 0 | 0 | 30 | 300 | 180 | 600 |
| Requirements specification | 0 | 0 | 780 | 120 | 570 | 600 | 0 | 240 | 0 | 0 | 0 | 240 | 120 | 810 | 1620 |
| Requirements review | 0 | 0 | 240 | 0 | 90 | 450 | 0 | 0 | 0 | 0 | 0 | 0 | 0 | 90 | 690 |
| Requirements adaptation | 0 | 0 | 0 | 0 | 30 | 60 | 0 | 0 | 120 | 0 | 0 | 0 | 0 | 30 | 180 |
| Architecture modeling | 0 | 0 | 240 | 270 | 565 | 0 | 0 | 405 | 0 | 180 | 0 | 30 | 450 | 970 | 270 |
| Architecture review | 0 | 0 | 360 | 0 | 0 | 120 | 0 | 0 | 0 | 0 | 0 | 0 | 90 | 0 | 480 |
| Architecture change | 0 | 0 | 0 | 0 | 0 | 0 | 0 | 0 | 0 | 420 | 0 | 0 | 420 | 0 | 0 |
| New statemate modeling | 0 | 0 | 0 | 720 | 0 | 0 | 660 | 1080 | 1620 | 765 | 810 | 0 | 2145 | 1890 | 1620 |
| Statemate review | 0 | 0 | 0 | 0 | 0 | 240 | 90 | 180 | 465 | 0 | 225 | 0 | 90 | 315 | 705 |
| Statemate change | 0 | 0 | 0 | 0 | 0 | 0 | 180 | 500 | 0 | 15 | 0 | 1200 | 195 | 500 | 1200 |
| Fault removal from statemate | 0 | 0 | 0 | 0 | 0 | 0 | 135 | 0 | 120 | 0 | 0 | 0 | 135 | 0 | 120 |
| Statemate optimization | 0 | 0 | 0 | 0 | 0 | 0 | 90 | 0 | 0 | 0 | 345 | 60 | 90 | 345 | 60 |
| Panel development | 0 | 0 | 0 | 0 | 240 | 0 | 540 | 0 | 0 | 0 | 0 | 0 | 540 | 240 | 0 |
| Panel change | 0 | 0 | 0 | 0 | 0 | 0 | 0 | 135 | 0 | 210 | 0 | 210 | 210 | 135 | 210 |
| System test | 0 | 0 | 0 | 0 | 0 | 480 | 180 | 60 | 0 | 420 | 530 | 450 | 600 | 590 | 930 |
| Integration | 0 | 0 | 0 | 0 | 0 | 0 | 0 | 0 | 0 | 0 | 0 | 330 | 0 | 0 | 330 |
| **Total effort** | 0 | 0 | 2220 | 1590 | 1915 | 2220 | 1965 | 2660 | 2385 | 2010 | 1910 | 2580 | 5475 | 6485 | 9405 |

**Table 1: Effort during the first iteration.**

The second, third, fourth, and five rows show the effort per week per group (EG1: emergent group 1, EG2: emergent group 2, V: V group) in minutes. Finally, the sixth row shows the total (i.e., for the first iteration) effort per activity. The effort distribution for other iterations looks similar. The table 1 shows that the effort of the V group is significantly larger than the effort of the emergent groups. In order to be able to evaluate product quality, we collected both internal and external metrics (see Table 2).

| Metrics type | Metric | EG 1 | EG 2 | VG |
|---|---|---|---|---|
| Indirect metrics | Number of activity charts | 2 | 4 | 5 |
|  | Number of state charts | 8 | 5 | 3 |
|  | Number of states | 56 | 56 | 59 |
|  | Number of state transitions | 83 | 96 | 107 |
|  | Are the state models executable | yes | yes | yes |
|  | Non-determinism | no | no | no |
|  | Data Dictionary (0-5) | 5 | 5 | 4, DOOR_X_OPEN undef |
|  | Clarity (0-5) | 4, as the architecture can be improved in the way that Chart EVENT CONTROL can be removed | 4,5; as it would be more clear, than to remove the parallelism in charts | 3, as many jumps are used |
|  | Architecture (0-5) | 4, the interfaces are well defined, but the presentation form can be improved. | 5, clear separation between hardware and software, the interfaces are well defined and presented | 3, the interfaces are not well defined |
| Direct metrics | Incorrectly implemented critical features | no | no | The bus can drive with open doors |
|  |  |  |  | The bus is not driving. If a button to open the door is pressed, the control system does not open the door. |
|  |  |  |  | If an accident happens, the bus does stop and does not open the doors. |
|  | Incorrectly implemented non-critical features | The doors do not have the button to open the door by passenger. | The button showing that the door is open takes the status "off" before the door is closed. | no |
|  |  | Driver light takes the status "on" before the door is completely open. | no | no |
|  |  | If outside is dark and the driver light switch has the status "off", the driver light is on. | no | no |

**Table 2: Data collected to evaluate product quality during first iteration.**

The first row in the table shows the type of the data collected. The second row shows the collected metric itself, for example, number of uncorrected implemented features. This number was defined as follows: first, we derived a standard test case set from the requirements specification. Second, we tested the delivered panels with respect to the set. Third, based on the knowledge about failed test cases, we identified features not correctly implemented. Furthermore, we separated the wrongly implemented features into critical and non-critical. In the first iteration, 18 different features should be implemented. In order to be able to focus on the critical features, we informed the groups (both emergent and V group) that in the first iteration, we would evaluate the quality only based on the critical features. Table 2 shows that the number of critical features incorrectly implemented by V group is larger than this number implemented by the emergent group. Consequently, the quality of the product implemented by the emergent group is higher that the quality of the product implemented by the V group. We assessed project team satisfaction by asking the students participating in the study about their satisfaction. The students evaluated their satisfaction based on the scale: high (=2), ok (=1), low (=0). For each group, we built a middle value per satisfaction aspect. This middle value is shown in Diagram 1. The diagram shows that the satisfaction with the work and with the task fulfillment is the same in the sub-group.

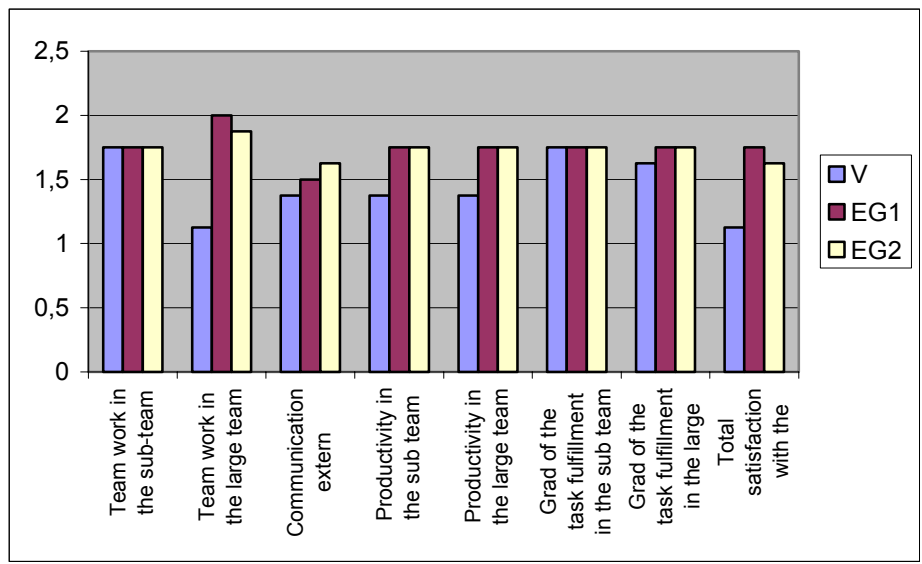

**Diagram 1: Team satisfaction**

All other satisfaction aspects are evaluated higher by emergent groups than by the V group. The results indicate that the EPAc method contributes to a more productive development of products, higher quality, and higher project team satisfaction.

## 5. Related Work

The approaches for acquisition of project-specific processes proposed in the literature can be divided into two types: top-down and bottom-up approaches. The top-down approaches can be further separated into three types: rule-based tailoring [13], constraint-based selection [16], and parameterized-based approaches. The bottom-up approaches provided in the literature can be classified in informal [2] and formal approaches [1].

The EPT method is neither an absolute bottom-up approach nor a top-down approach. It is an approach that uses top-down tailoring at the start of the project to reuse knowledge gained in past similar projects, and refines the top-down tailored process based on the performed tasks. This two-step tailoring approach allows avoiding deficiencies of the bottom-up and top-down approaches.

## 6. Summary and Future Work

Efficient development of qualitative systems requires suitable project-specific processes. As projects have different contexts and goals, tailoring methods are needed that allow adapting the generic processes to the project-specific needs. The tailoring approaches used in practice (e.g., the tailoring approach proposed by the V model [11]) usually involve checking conditions and removing objects of the base model. One difficulty of such tailoring is the identification of the regression process modification to be performed if an object is removed. Furthermore, process tailoring often requires not only removal of process objects, but also replacement or addition of new objects. Traditional tailoring methods do not define how to deal with such kinds of process modifications.
To tackle the problem, we provide the Emergent Process Acquisition (EPAc) method. This method uses the domain-specific process line for top-down tailoring and refines the tailored process based on the process activities performed in the first process iteration. In this way, the initial variant of the emergent process is built. This paper presented the emergent process acquisition method and the empirical experience gained with the method in the context of a study. The study shows that the emergent tailoring method significantly contributes to more efficient development of higher-quality systems.
One issue for future fwork is the validation of the EPAc method in a real context.

## 7. Acknowledgement

The authors would like to thank Alexander Raschke, Matthias Schneiderhan, Ramin Tavakoli Kolagari, and Frank Houdek for very helpful support during validation. Furthermore, our thanks go to the following people who provided valuable input in several discussions: Dieter Rombach, Kurt Schneider, Michael Stupperich. The authors would also like to thank Sonnhild Namingha from the Fraunhofer Institute for

Experimental Software Engineering (IESE) for reviewing the first version of the article.